\documentclass[aps,pre,reprint,superscriptaddress,showkeys]{revtex4-2}

\usepackage{graphicx}\usepackage{dcolumn}\usepackage{bm}\usepackage{amsfonts}
\usepackage{amsmath}
\usepackage{amssymb}
\usepackage{amstext}
\usepackage{latexsym}
\usepackage{textcomp}
\usepackage[separate-uncertainty=true]{siunitx}
\usepackage[usenames,dvipsnames]{xcolor}

\newcommand{\modified}[1]{{#1}}

\begin{document}

\title{Pattern Formation in Slot-Die Coating}
\author{Maren Kasischke}
\affiliation{Ruhr-Universit\"at Bochum, Chair of Applied Laser Technology,
    Universit\"atsstra\ss{}e~150, 44801 Bochum, Germany}
\author{Simon Hartmann}
\email{s.hartmann@wwu.de}
\thanks{ORCID: 0000-0002-3127-136X}
\affiliation{Institut f\"ur Theoretische Physik, Universit\"at M\"unster, Wilhelm-Klemm-Str.\ 9, D-48149, M\"unster, Germany}
\author{Kevin Niermann}
\affiliation{Ruhr-Universit\"at Bochum, Chair of Applied Laser Technology,
    Universit\"atsstra\ss{}e~150, 44801 Bochum, Germany}
\author{Denis Kostyrin}
\affiliation{Ruhr-Universit\"at Bochum, Chair of Applied Laser Technology,
    Universit\"atsstra\ss{}e~150, 44801 Bochum, Germany}
\author{Uwe Thiele}
\email{u.thiele@uni-muenster.de}
\homepage{http://www.uwethiele.de}
\thanks{ORCID: 0000-0001-7989-9271}
\affiliation{Institut f\"ur Theoretische Physik, Universit\"at M\"unster, Wilhelm-Klemm-Str.\ 9, D-48149, M\"unster, Germany}
\affiliation{Center for Nonlinear Science (CeNoS), Westf{\"a}lische Wilhelms-Universit\"at M\"unster, Corrensstr.\ 2, 48149 M\"unster, Germany}

\author{Evgeny L. Gurevich}
\email{gurevich@lat.rub.de}
\thanks{ORCID: 0000-0001-9451-8983}
\affiliation{Ruhr-Universit\"at Bochum, Chair of Applied Laser Technology,
    Universit\"atsstra\ss{}e~150, 44801 Bochum, Germany}
\affiliation{University of Applied Sciences M\"unster, Laser Center (LFM), Stegerwaldstra\ss{}e 39, 48565, Steinfurt, Germany}
\date{\today}

\begin{abstract}
    We experimentally study the occurrence of pattern formation during the slot-die coating of low-viscosity nearly Newtonian liquids onto Polyethylenterephthalat (PET)-substrates. In particular, it is demonstrated that with increase of the coating speed a homogeneous coating becomes unstable with respect to periodic stripe patterns. Thereby, depending on the liquid viscosity, the stripes can be oriented parallel or perpendicular with respect to the coating direction. Mixed states do also occur. The spatial period of perpendicular [parallel] stripes increases [decrease] with the coating speed. The dependence of the effect on various control parameters is investigated.  Finally, a simple theoretical model based on the hydrodynamics of thin films of partially wetting liquids is analyzed. Comparing the results to the experiments, conclusions are drawn regarding the acting instability and pattern formation mechanisms.
\end{abstract}


\maketitle

\section{Introduction}

The coating of flexible substrates with thin homogeneous layers is an important technological process relevant for many areas of modern life and various fields of technology such as, e.g., analog photography, surface chemistry~\cite{chang2013}, flexible electronics~\cite{sandstrom2012}, and organic photovoltaics~\cite{krebs2009, Jakubka2013}. Driven by technological needs, scientific investigations of the coating process, e.g., for slot-die or blade coating are usually focused on specifying the \textit{coating window}, i.e., the range of process parameters where a \textit{homogeneous} defect-free coating of constant thickness is reproducibly achieved~\cite{Gutoff2006, DiLH2016aj}. Depending on fluid and substrate properties and process parameters, outside this range instabilities occur that result in coating defects and patterned coatings~\cite{Bhamidipati2012,Raupp2018}. The observed patterns include stripes parallel and orthogonal to the coating direction similar to patterns observed in other deposition processes~\cite{Thie2014acis,Lars2014aj}, including dip coating~\cite{BoLe2010sm,DoGu2013e,BDLL2013cep}, Langmuir-Blodgett transfer of surfactant layers onto a moving plate~\cite{GlCF2000n,LKGF2012s,KoTh2014n} and evaporative dewetting (aka ``coffee-ring effect'')  where a three-phase contact line driven by evaporation recedes on a solid substrate~\cite{HaLi2012acie,FrAT2012sm}. The focus of the present work is an analysis of the occurring patterns and their sequence.

Sometimes, e.g., in the production of organic solar cells, initial coating steps that produce uniform deposited layers are followed by subsequent technological steps producing periodic patterns. This is, e.g., achieved by laser ablation of the layers~\cite{SolarCells}. Such post-processing patterning steps could be omitted if direct patterning by coating can be controlled in a similar way as control of patterning by Langmuir-Blodgett transfer and evaporative dewetting is used in applications~\cite{LZMW2004am,MaEr2018acis}. A prerequisite is a detailed investigation and understanding of the nonequilibrium pattern formation that occurs outside the coating window. Here, we investigate the transition from homogeneous to patterned coating and the characteristics of the patterning process in dependence of the most important control parameters for slot-die coating of low-viscosity nearly Newtonian liquids onto Polyethylenterephthalat (PET)-substrates. The focus is on slot-die coating as, similar to the closely related blade coating, it allows for rapid processing of large surface areas at standard atmospheric conditions, i.e., it can be incorporated into roll-to-roll processing.

The self-organized patterns emerging during coating are usually treated as unwelcome defects and are therefore barely studied in the literature. Normally, experimental investigations that consider the onset of coating instabilities, do not focus on the resulting pattern formation. Bhamidipati et al.~\cite{Bhamidipati2012} study patterning in slot-die coating induced by air entrainment in viscous non-Newtonian fluids. They observe the formation of periodic stripes and regular arrangements of bubbles in shear-thinning, non-Newtonian liquids with relatively high viscosity. Raupp et al.~\cite{Raupp2018} observe breakup of a homogeneous film and formation of parallel stripes in a certain range of low film thicknesses and coating velocities. In both cases, the orientation of the stripes is parallel to the coating direction, i.e., the direction of slot-die movement. With other words, the stripes are perpendicular to the slot in the coating device. These stripes are referred to as ``rivulets'' while a weaker thickness modulation is referred to as ``ribbing''.

The ranges of capillary numbers and of the distance between slot-die and substrate (gap height) that correspond to the coating window are also investigated for shear-thinning liquids~\cite{Schmitt}, where stripes oriented parallel (``barring'') and perpendicular to the coating direction are observed. Mixed patterns are also possible. A review of operating limits in slot-die coating is given in Ref.~\cite{DiLH2016aj}, however, little information is given on the patterning phenomena outside the coating window. Lin et al.~\cite{lin2010} employ numerical simulations to predict the parameters for homogeneous coating and find different types of coating defects outside the coating window. At high coating speeds structures referred to as ``breaklines'' (lines of varying shape and width) coupled with ``dripping'' (drops spraying from the die) are observed, while ribbing and air entrainment limit the coating window if the flow rate is too low for the chosen coating velocity.

Only stripes parallel to the coating direction are observed in Ref.~\cite{lin2010}. In terms of coating speeds the predicted coating window is larger than the experimentally determined one. Kang et al.~\cite{kang2014} find that the coating velocity has a dominant effect on both, the thickness and the width of the stripes. The thickness is proportional to the velocity while the width is inversely proportional to it. Khandavalli et al.~\cite{khandavalli2016} study the impact of shear-thickening of the coating fluid on the stability of slot-die coating. The slot velocity for the onset of coating defect occurrence and the type of coating defect are examined at different coating parameters such as flow rate and coating gap height. They find that shear-thickening has a negative impact on the coating window and the intensity of the ribbing instability is found to increase with shear-thickening magnitude.

Note that certain specific characteristics of the described patterning processes in slot-die coating are shared by the other above mentioned coating and deposition techniques dip coating, Langmuir-Blodgett transfer and evaporative dewetting~\cite{Lars2014aj,HaLi2012acie,Thie2014acis,LZMW2004am,MaEr2018acis}. These processes are normally employed to coat small areas. A recent detailed analysis of patterns obtained in evaporative dewetting is given in Ref.~\cite{JEZT2018l}. In the case of  the well studied ``passive systems'' the contact line velocity is selected by the system and not externally imposed as in ``active systems'' like slot-die coating and many other coating processes~\cite{Thie2014acis}. The latter are studied for simple and some complex liquids, e.g., for spin coating~\cite{PBPG2013sm} and dip coating (also called ``dragged plate geometry'')~\cite{MRRQ2011jcis,LKGF2012s}, however, they are normally not suitable for large-scale industrial application. However, besides plenty of experimental results, for these systems modeling  of pattern formation is also quite advanced. Thin-film models, i.e., hydrodynamic long-wave models and amended Cahn-Hilliard models, are used to analyze the bifurcation behavior of pattern formation processes in evaporative dewetting~\cite{FrAT2012sm}, dip coating~\cite{DoGu2013e,DeDG2016epje,TWGT2019prf} and Langmuir-Blodgett transfer~\cite{KGFT2012njp,KoTh2014n}. Normally, this is done for one-dimensional substrates, but first bifurcation results also exist in the two-dimensional case (see~\cite{LMTG2020pd} and section~8.2 of~\cite{EGUW2019springer}). Comparison of results for slot-die coating with other coating and deposition techniques will allow one to identify universal features of the occurring instabilities and patterning processes.

Here, we focus on an experimental study of pattern formation outside the coating window for slot-die coating. We demonstrate the existence of stripes oriented parallel and perpendicular with respect to the coating direction. Mixed patterns do also occur. The conditions of onset of pattern formation are discussed as well as the dependencies of the properties of the patterns on the experimental control parameters. The experimental findings are compared to theoretical results obtained using a thin-film model for slot-die coating that takes capillarity and wettability into account as well as parameters of the coating process like coating velocity and gap height.

Our work is structured as follows. In the following section~\ref{sec:exp} the experimental setup is introduced, while section~\ref{sec:results} presents the corresponding results. Therein, the first subsection focuses on general coating scenarios while another subsection discusses the influence of the coating gap height. Section~\ref{sec:theo} briefly introduces the employed model and describes the obtained results. Finally, we present a general discussion and conclusion in section~\ref{sec:diss}.

\section{Experimental Setup}\label{sec:exp}
\subsection{Coating System}
\begin{figure*}
    \centering
    \includegraphics[width=17.8cm]{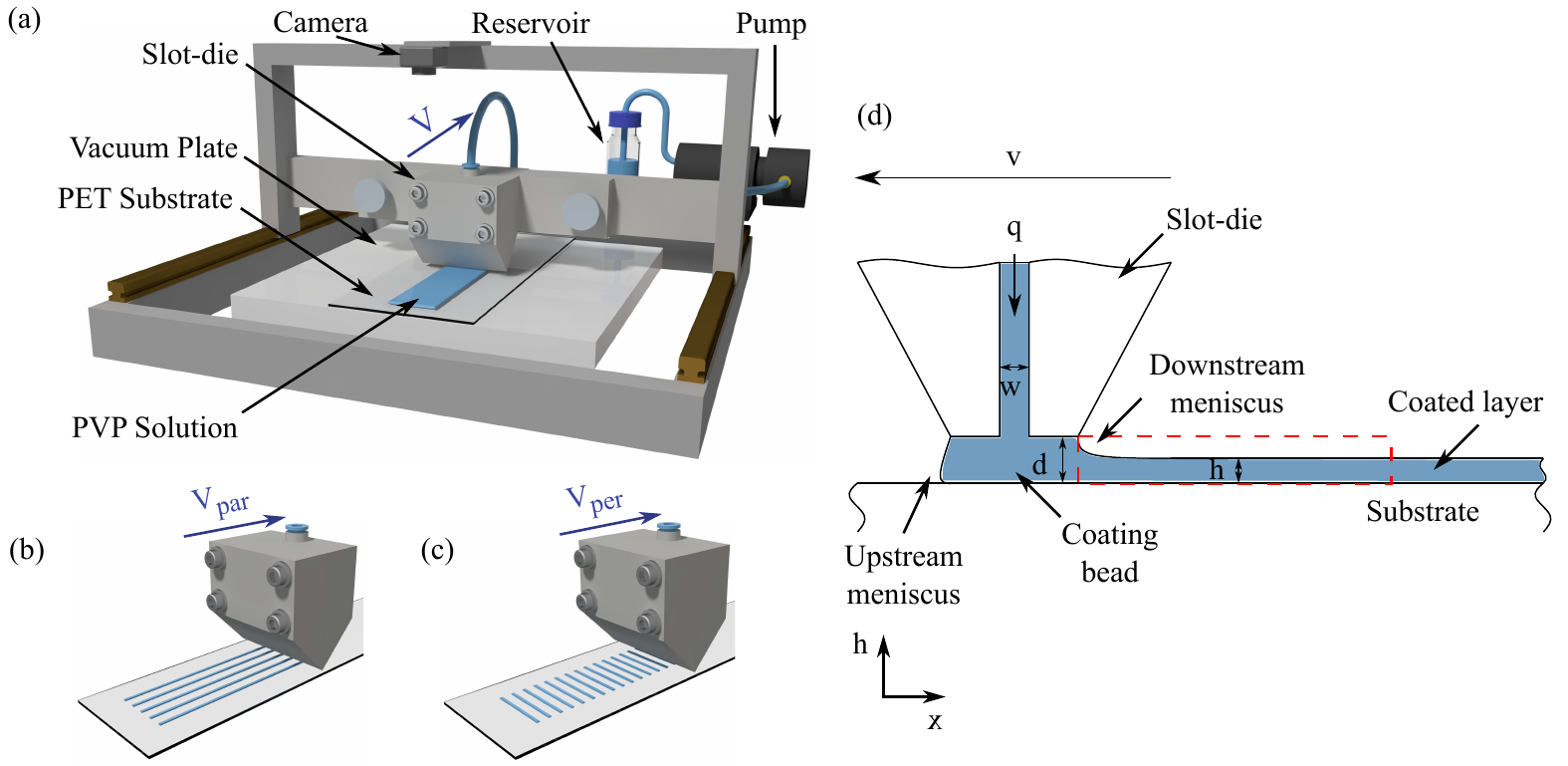}
    \caption{(a) Experimental set up of slot-die coater with video camera for capturing the coated thin-film and formation of stripes with (b) parallel and (c) perpendicular orientation to the coating direction. (d) Schematic representation showing a cross section of the essential part of the slot-die, the liquid inflow and coated layer.}\label{SetUp}
\end{figure*}

To investigate the pattern formation occurring in slot-die coating we employ as 
experimental model system the coating of a polyethylenterephthalat 
(PET)-substrate (P\"utz GmbH, Germany) with a polyvinylpyrrolidon (PVP) K90 
solution (VWR International GmbH, Germany; \modified{CAS: 9003-39-8, molecular weight: $\SI{360 000}{g/mol}$, K value range: 90--103}) in ethanol. PET-substrates are very promising for 
low-cost applications in flexible electronics and in solar cells. The 
PVP-Ethanol mixtures allow for a rather flexible adjustment of the physical 
characteristics of the coated liquid by varying the PVP content. 
Fig.~\ref{SetUp} presents the experimental setup and a schematic representation 
of a cross section of the essential part of the slot-die, the liquid inflow and 
coated layer. The coater (Easycoater, Coatema Machinery GmbH, Germany) is a 
discontinuous coater for single piece coating.
Compared to roll coating used in most other studies, the discontinuous coater has the advantage that
no part of the substrate is re-used as happens when the same part of the roll returns after each full turn. In contrast, in each slot-die experiment reproducibly a fresh unused substrate is coated. Furthermore, it offers a fast and simple method to study slot-die coating on a small scale with a low waste of substrates and chemicals.

\subsection{Coating Layers}
\label{sec:exp-coat}
The width of the coating area (transversal size) corresponds to the length of the slot in the coating die and is approximately \SI{8}{cm}. The total length of the coating area (longitudinal size) is about \SI{25}{cm}. However, only the central third of the total length is used for further analysis, i.e, a length of approximately \SI{8}{cm} length. The remaining longitudinal range shows transient behavior due to inevitable acceleration at the start and deceleration at the end of a coating run. 

Between different experiments the coating velocity is varied from $v_C=\SI{0.4}{\meter\per\minute}$ to $v_C=\SI{4.4}{\meter\per\minute}$ in steps of $\SI{0.2}{\meter\per\minute}$. The coating gap height is varied by increasing it from $d=\SI{220}{\micro\m}$ in steps of $\SI{30}{\micro\m}$ till $d=\SI{430}{\micro\m}$ and then in larger steps ($\SI{60}{\micro\m}$ and $\SI{90}{\micro\m}$) till $d=\SI{730}{\micro\m}$. Finally, an experiment at $d=\SI{1210}{\micro\m}$ is performed, bringing the overall increment to $\SI{990}{\micro\m}$. Unless stated otherwise the coating gap height is fixed at $d=\SI{220}{\micro\m}$. The width of the slot channel is $w=\SI{100}{\micro\m}$ in all experiments.

Note that along total length of the coating area of $\SI{25}{\cm}$ the coating gap has a small variance of approximately $\SI{60}{\micro\m}$: For instance, at the start point the coating gap height is $d=\SI{191}{\micro\m}$ while at the end point it is $d=\SI{254}{\micro\m}$. We refer to such a run as having gap height $d=\SI{220}{\micro\m}$ but keep in mind that it is actually $d\approx\SI{220\pm 30}{\micro\m}$. This implies that over the relevant central third of the length of the coating area we have about 10\% variation for the $d=\SI{220}{\micro\m}$ gap and about 2\% variation for the $d=\SI{1210}{\micro\m}$ gap. In contrast, there is no measurable variation of the coating gap along the width of the coating area, i.e, the die is very well aligned parallel to the substrate.

A variation in the coating gap is reported to be an important parameter, which can influence the pattern formation in some systems like for confined dip coating~\cite{kim_nam_2017}. This raises the question whether the described small variation in the coating gap along the length of the coating area can influence the pattern formation in our experiments. In section~\ref{sec:gap} we show that this imperfection does not influence the pattern formation process in the considered parameter range.

The pattern formation is analyzed during the coating process by means of the CCD camera (VRmC PRO, VRmagic GmbH, Germany) installed directly on the die. Later, after finishing the coating process, the resulting pattern is analyzed via high-resolution photographs (OM-D E-M10 Mark III with macro objective, Olympus, Japan). Comparing the results of the two methods, we conclude that the already deposited pattern does not noticeably change, e.g., due to ethanol evaporation, during the further coating process or shortly thereafter. Blue ink is added to the ethanol-PVP mixtures in a small concentration to increase the contrast to facilitate optical imaging. No differences in the patterns and their characteristics, such as contact angle and viscosity, are observed in runs with and without ink. Therefore, the influence of the ink on the properties of the mixture can be neglected.

\subsection{Coating fluid characterization}
\modified{We assume that the employed 5-9\%  PVP solutions can be considered to be nearly Newtonian: Although Ref.~\cite{goodwin2000} uses a 12\% PVP solution as an example of a shear-thinning liquid giving in their Fig.~1.5 the behaviour for a shear rate ranging from 0 to \SI{600}{s^{-1}}, the power-law behaviour is rather weak. While their fit indicates a power of 0.86, a linear (i.e., Newtonian) fit shows only a small deviation from their curve (not shown).}

\begin{figure}
    \centering
    \includegraphics[width=8.6cm]{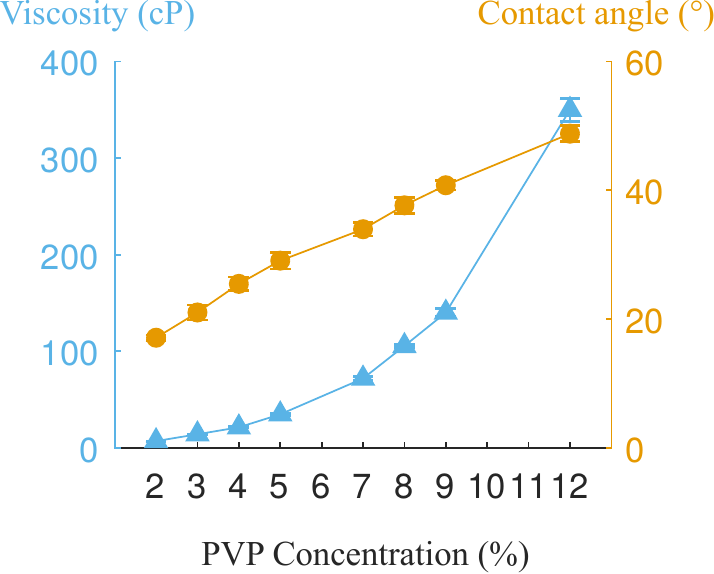}
    \caption{Viscosity of the ethanol-PVP mixture (left vertical axis, triangles) and advancing contact angles in degrees (right vertical axis, circles) as functions of the mass concentration of PVP in the solution.}\label{CAvisc}
\end{figure}

It is important to note, that not only the viscosity (as expected and intended) but also the contact angle of the Ethanol-PVP mixture on the PET Substrate is found to depend quite strongly on the polymer concentration. We use a setup built in-house and an ImageJ plugin~\cite{ImageJ} to fit the shape of static sessile droplets after inflation with a syringe and measure the advancing contact angle from the video data. The resulting dependencies of the contact angle and the viscosity on PVP concentration are presented in Fig.~\ref{CAvisc}. The measured viscosity in cP corresponds to 1\,cP=10$^{-3}\,$Pa\,s. Changing the mass concentration of PVP from 2\% to 12\%, the viscosity increases 50-fold while the contact angle still triples from about $17^\circ$ to about $50^\circ$ what cannot be considered a small change. This strong change must be due to dependencies of the liquid-gas and liquid-solid interface tensions on PVP concentration. The  liquid-gas interface tension of PVP-in-ethanol solutions is known to slightly decrease from $\sigma_\text{PVP}\approx\SI{24}{\milli\newton\per\meter}$ to $\SI{23}{\milli\newton\per\meter}$~\cite{taghizadeh2017} for an increase in concentration from 0\% to 20\%. As this cannot explain the observed change in contact angle even qualitatively, the liquid-solid interface tension must increase with PVP concentration such that the difference of solid-gas and solid-liquid interface tensions decreases by about 1/3.

\section{Experimental Results}\label{sec:results}
\subsection{Coating scenarios}\label{sec:scenarios}
\begin{figure}[h]
    \centering
    \includegraphics[width=8.6cm]{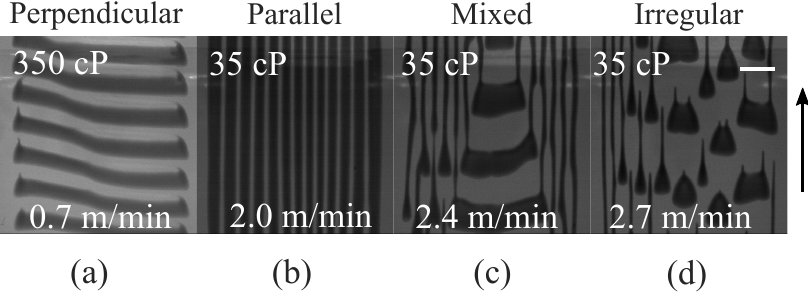}
    \caption{Images of (a) stripes oriented perpendicular to the direction of coating, (b) stripes oriented parallel to the direction of coating (c) mixture of stripes perpendicular and parallel to the direction of coating and (d) irregular pattern. The viscosity and coating velocity are given in the images. Dark areas represent the coating, while bright areas correspond to the uncoated substrate. The scale bar is 10 mm. The arrow indicates the coating direction.}\label{example}
\end{figure}

Depending on the parameters, homogeneous as well as inhomogeneous coatings are deposited on the substrate. In general, four types of coating patterns are observed beside the uniform coating. They are \begin{enumerate}
    \item parallel stripes oriented perpendicular to the direction of coating [Fig.~\ref{example}~(a)]; \item parallel stripes oriented parallel to the direction of coating [Fig.~\ref{example}~(b)]; \item irregular mixed patterns, corresponding to a combination of the two aforementioned stripe patterns [Fig.~\ref{example}~(c)];
    \item irregular patterns, corresponding to a set of rather randomly located dots [Fig.~\ref{example}~(d)].
\end{enumerate}

The mixed patterns consist of areas covered by stripes of different orientation, i.e., the stripes in different substrate regions are perpendicular to each other. This mixed-pattern regime is observed in a parameter range at the border between the parameter regions where parallel and perpendicular stripes occur, respectively. Note that sometimes the perpendicular stripes are slightly tilted with respect to the direction of coating. Because the tilt angle differs between the runs [cf.~e.g., Fig.~\ref{example} (a) and~(c)] we believe the effect is not due to imperfections in the gap height, but rather represents defects (phase slips) in the line pattern.

The pattern topology and quantitative characteristics like typical pattern length scales are influenced by several processing parameters (e.g., the coating velocity $v$, the coating gap height $d$, the viscosity of the liquid mixture $\eta$, its surface tension $\sigma$, and the liquid flow rate $q$). In the following, we discuss the main influences. First, we keep the flow rate $q=\SI{4.8\pm0.2}{\milli \litre / \minute}$ (or \SI{1}{\milli\metre^2/\second} per unit width) and the gap height $d=\SI{220}{\micro\m}$ fixed. The mentioned small drift in the gap height along the coating direction is less than about \SI{60}{\micro\m} over the full die length of \SI{25}{cm} corresponding to an average slope of less than \SI{2e-4}{rad}. Such small imperfections of the experimental setup cannot be avoided due to the adjustment of the vacuum plate, which holds the substrate, and the linear guide of the slot-die over the coating length of \SI{25}{cm}. Next, we investigate the coating patterns in dependence of two main technology-relevant parameters: coating velocity $v$ and viscosity $\eta$. However, we emphasize that the variation in viscosity is achieved by changing the chemical composition of the mixture and also influences the surface tension $\sigma$ and the equilibrium contact angle $\theta$.

\begin{figure*}
    \centering
    \includegraphics[width=17.8cm]{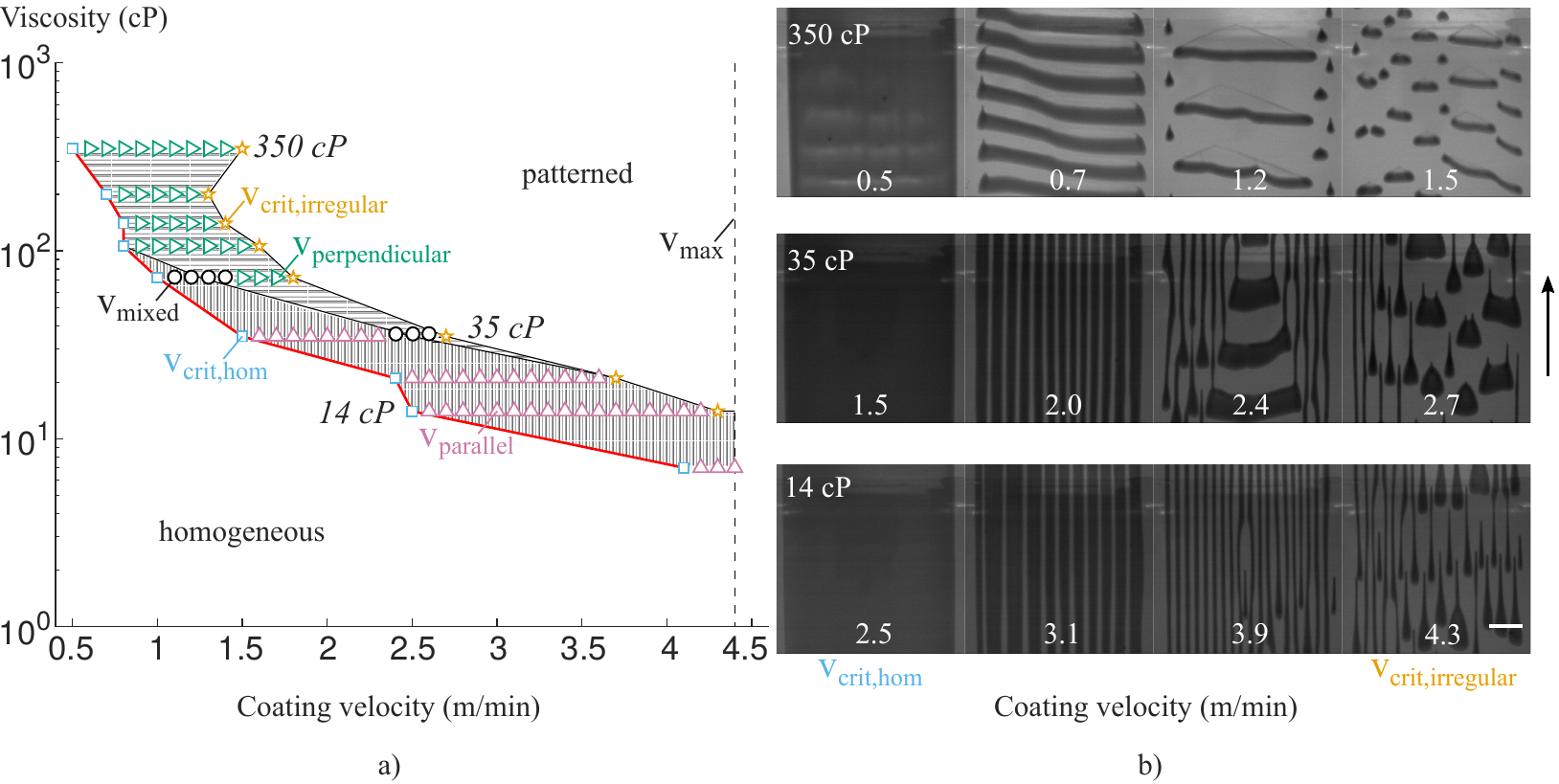}
    \caption{(a) The morphological phase diagram presents the parameter regions corresponding to the various coating patterns in the plane spanned by coating velocity $v$ and viscosity $\eta$. The region corresponding to homogeneous coating is at velocities $v \leq v_\mathrm{crit,hom}$ (marked by squares and the red line), irregular patterns occur at velocities $v \geq v_\mathrm{crit,irregular}$ (marked by stars). Parallel (marked as $\vartriangle$) and perpendicular (marked as $\triangleright$) stripes are found at $v_\mathrm{parallel}$ and $v_\mathrm{perpendicular}$, respectively, and mixed patterns at $v_\mathrm{mixed}$. Vertically and horizontally hatched regions mark the occurrence of parallel and perpendicular stripe patterns, respectively. (b) Transitions from homogeneous coating to stripes upon increasing the coating velocity for three different viscosities. The coating direction in the figures is from the bottom to the top and the scale bar is 10 mm.}\label{map}
\end{figure*}

Fig.~\ref{map}~(a) shows a morphological phase diagram indicating where in the parameter plane spanned by coating velocity and viscosity the various coating types and patterns occur. In general, at low velocities and low viscosities the coating is homogeneous, i.e., the substrate is covered by a liquid layer of constant thickness. This parameter region is also known as the coating window. With increasing velocity a transition occurs from the homogeneous coating to coating defects due to instability of the meniscus. Thus, the red curve marked by blue open squares corresponds to the critical coating velocity $v_\mathrm{crit,hom}$ for each viscosity. At high velocities and high viscosities above $v_\mathrm{crit,irregular}$ irregular patterns prevail. The transition scenario found when going from one to the other limiting case depends on the liquid viscosity. Images in Fig.~\ref{map}~(b) show typical transitions in the observed patterns at fixed high, intermediate, and low viscosity.

Inspecting the first row in Fig.~\ref{map}~(b), one observes that at high viscosities [or high contact angles, cf.~Fig.~\ref{CAvisc}], with increasing $v$ the homogeneous coating (e.g., at $v_\mathrm{crit,hom}=\SI{0.5}{\meter\per\minute}$) is replaced by perpendicular stripes (e.g., at $v=\SI{0.7}{\meter\per\minute}$) that are slightly bended at the lateral sides of the coating area. The stripes are horizontal in the figure as the slot-die is horizontal with respect to the panels and moves vertically upwards across them (as indicated by the arrow). Upon further increase of $v$, the stripes break up by the subsequent introduction of phase slips (e.g., at $v=\SI{1.2}{\meter\per\minute}$). Ultimately, the pattern becomes quite irregular and consists partly of dots that can be slightly elongated in perpendicular direction (e.g., at $v_\mathrm{crit,irregular}=\SI{1.5}{\meter\per\minute}$).
In contrast, at low viscosity [bottom row in Fig.~\ref{map}~(b)], with increasing $v$ the homogeneous coating (e.g., at $v_\mathrm{crit,hom}=\SI{2.5}{\meter\per\minute}$) is replaced by parallel stripes (e.g., at $v=\SI{3.1}{\meter\per\minute}$). Upon further increase in $v$, the stripes first develop asymmetric bulges that at larger $v$ develop into defects where stripes split or end (e.g., at $v=\SI{3.9}{\meter\per\minute}$). Towards larger velocities stripes mostly end at bulges. Ultimately, only the bulges survive, showing short stripe-like appendices and the pattern becomes quite irregular (e.g., at $v_\mathrm{crit,irregular}=\SI{4.3}{\meter\per\minute}$). Overall, the transition to patterns occurs at larger velocities in the low viscosity case than in the high viscosity case [cf.~Fig.~\ref{map}~(a)].

\begin{figure*}
    \centering
    \includegraphics[width=17.8cm]{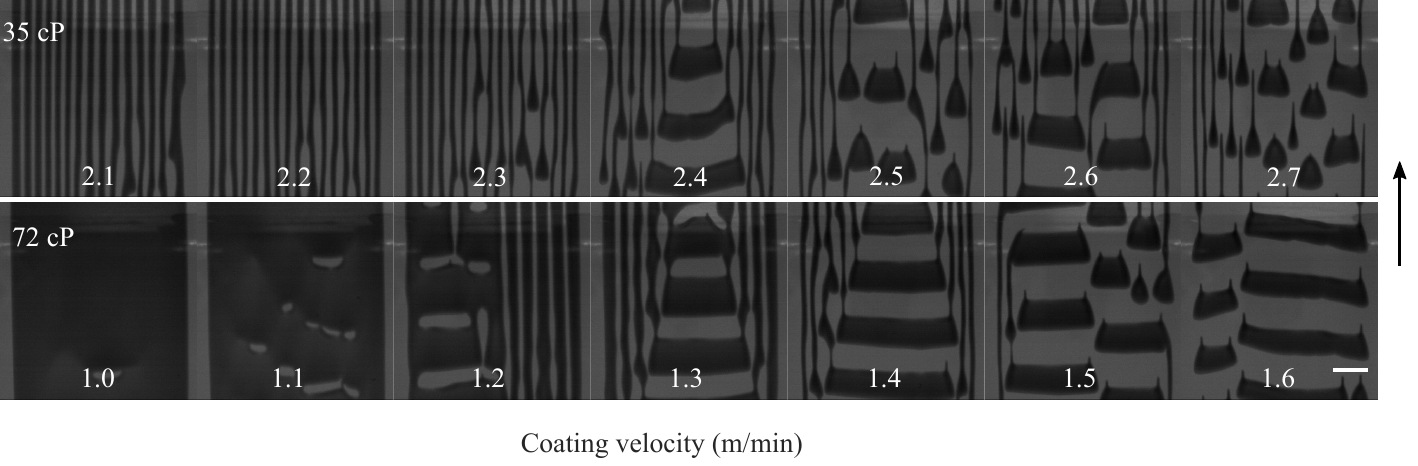}
    \caption{Typical transition scenario of coating patterns in dependence of the coating velocity $v$ at fixed intermediate viscosities $\eta=\SI{35}{cP}$ and $\eta=\SI{72}{cP}$. With increasing $v$ a transition occurs to coexisting perpendicular and parallel stripes and further to perpendicular stripe patterns with defects. Remaining details are as in Fig.~\ref{map}.}\label{HorVert}
\end{figure*}

As further discussed in section~\ref{sec:gap}
one needs a relatively large change in the gap height to qualitatively change the pattern. For instance, the stripes parallel to the coating direction are a robust state, they even remain when the gap height is changed by a factor of three: from $d\approx\SI{0.4}{mm}$ to $d\approx\SI{1.2}{mm}$. As the variation in $d$ along the length of the coating area is much smaller (less than $\pm10\%$ in all experiments, see section~\ref{sec:exp-coat}), this imperfection in the experimental setup has negligible influence on the pattern formation.

In a small velocities range at intermediate viscosity, one observes the simultaneous occurrence of perpendicular and parallel stripes in different regions of the substrate [see second row in Fig.~\ref{map}~(b)]. The transition scenario for intermediate viscosity, e.g.\ $\eta=\SI{35}{cP}$ and $\eta=\SI{72}{cP}$ at 5\% and 7\% of PVP, respectively, is shown in Fig.~\ref{HorVert}. At $\eta=\SI{35}{cP}$ the homogeneous coating first destabilizes to parallel stripes. With increasing velocity patches of perpendicular stripes appear within the pattern of parallel stripes. At $\eta=\SI{72}{cP}$, as the coating velocity is increased, the homogeneous coating is destabilized and first pieces of perpendicular empty stripes appear ($v=\SI{1.1}{\meter\per\minute}$). Further increasing $v$, the uncovered stripes grow in length resembling a perpendicular stripe pattern. However, in part of the coating area also regions of parallel stripes develop ($v=1.2\dots\SI{1.4}{\meter\per\minute}$). At intermediate velocities, areas filled with parallel and perpendicular stripes seem to coexist in every individual coating experiment. At higher velocities perpendicular stripes dominate before breaking up via phase slips as described above.

\begin{figure}
    \centering
    \includegraphics[width=8.6cm]{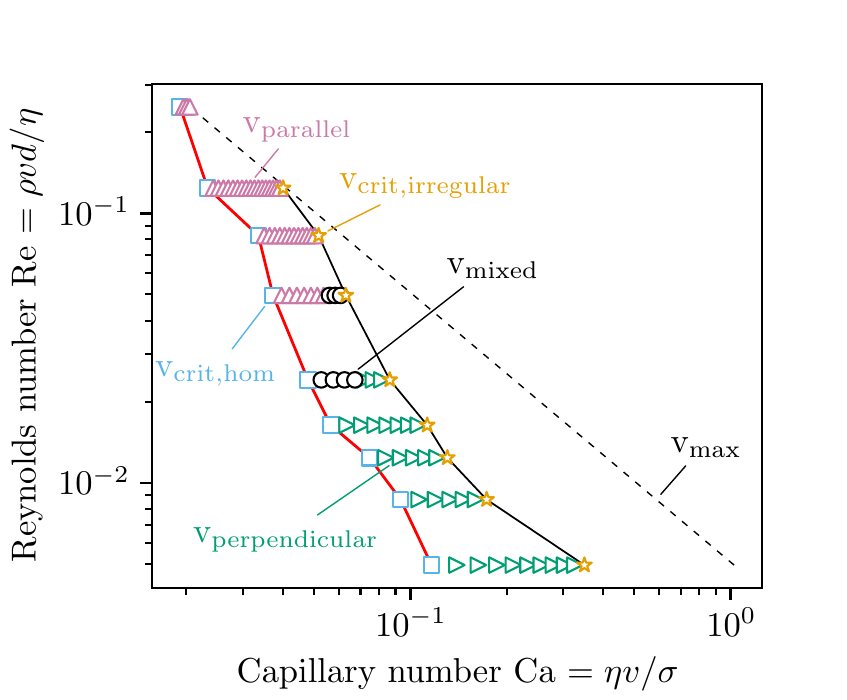}
    \caption{The data of the phase diagram in Fig.~\ref{map}~(a) shown in the plane spanned by the Capillary and Reynolds numbers. All colors and markers are as in Fig.~\ref{map}~(a).}\label{map_dimless}
\end{figure}

Further, we have converted the data of the phase diagram in Fig.~\ref{map}~(a) to dimensionless quantities, in order to allow for a more general comparison with other related experiments of coating instabilities. As typically found in the literature~\cite{CaKh2000aj, RoSC2006jnfm, MaCa2015aj}, we choose the Capillary number $\mathrm{Ca} = \eta v / \sigma$ and the Reynolds number $\mathrm{Re} = \rho v d / \eta$ to span the phase plane for the rescaled diagram in Fig.~\ref{map_dimless}. Here, we find that the observed instabilities occur in a laminar regime ($\mathrm{Re} \ll 1$) and both the Reynolds number and the Capillary number are involved in the selection of the pattern.

\begin{figure}
    \centering
    \includegraphics[width=8.6cm]{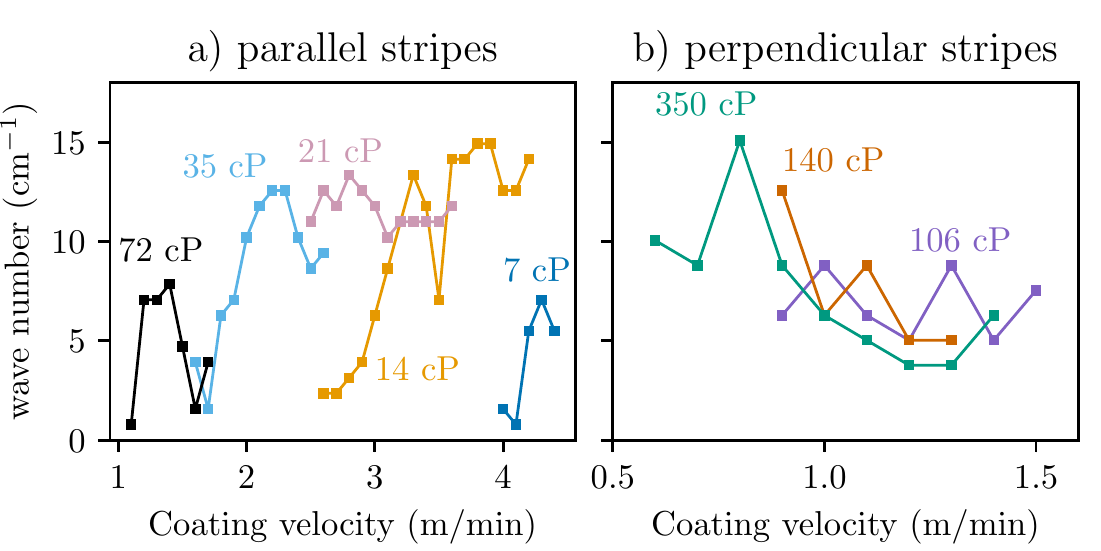}
    \caption{(a) The wavenumber of parallel stripes and (b) the wavenumber of perpendicular stripes in the central coating area are given in dependence of the coating velocity at different viscosities.}\label{PeriodPattern}
\end{figure}

For all performed experiments with a clear stripe pattern, the period of the stripes in the central (nontransient) coating area of approximately \SI{8}{cm} length is measured. The dependency of the wavenumber $k=2\pi N/L$ of the stripes on the coating velocity is presented in Fig.~\ref{PeriodPattern}~(a) and~(b) for the perpendicular and parallel stripes, respectively, where $L$ denotes the extent of the stripes. The perpendicular stripes observed at high viscosity (cf.~Fig.~\ref{map}) show (up to an outlier) a decreasing wavenumber (proportional to the inverse period). The behavior is different for the parallel stripes, where the wavenumber increases with the velocity.

\subsection{Influence on the coating gap height}\label{sec:gap}
Next we study the influence of the coating gap height by step-wise increasing it at constant coating velocity $v_\mathrm{crit,hom}$, which is the largest velocity for which a homogeneous coating is achieved for a given viscosity at the lowest considered gap height. This ensures that we are always within the parameter range where patterns occur. In this way we can also assess which influence small changes in the gap height have (like the variation related to the small imperfectness in the experimental setup discussed above). This procedure is followed for a number of different viscosities.

\begin{figure*}
    \centering
    \includegraphics[width=17.8cm]{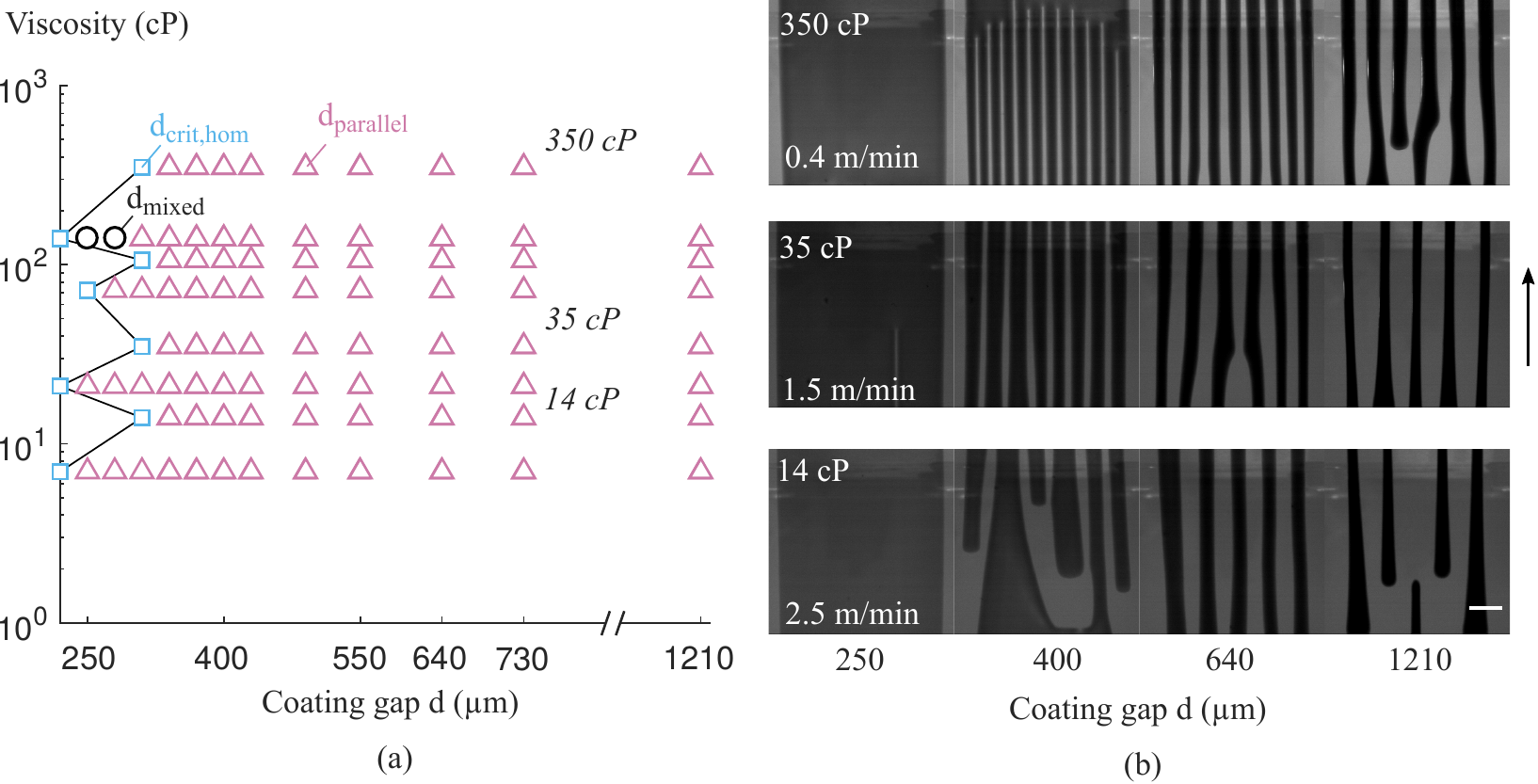}
    \caption{(a) Morphological phase diagram for the parameters corresponding to the transition from homogeneous coating to stripes upon increasing the coating gap height $d$, starting from $d=\SI{220}{\micro\m}$. The velocity used for each viscosity is $v_\mathrm{crit,hom}$ (see Fig.~\ref{map}). (b) Transition of patterns with increasing $d$ at different fixed viscosities. Remaining details are as in Fig.~\ref{map}}\label{gap}
\end{figure*}

Figure~\ref{gap}~(a) gives an overview of the conducted experiments and the resulting patterns. When the coating gap height is increased above a critical value $d_\mathrm{crit,hom}$, a transverse destabilization of the meniscus is observed and the homogeneous coating breaks up into stripes parallel to the coating direction. Unlike the observations made above when increasing the coating velocity, here the orientation of the stripes remains unchanged when increasing the viscosity. Also, there is no pronounced influence of viscosity on the onset of pattern formation, i.e., on $d_\mathrm{crit,hom}$, see Fig.~\ref{gap}~(a).

With increasing coating gap height the distance between stripes increases and their width decreases. This occurs at all tested viscosities, see the example patterns in Fig.~\ref{gap}~(b).

\begin{figure*}
    \centering
    \includegraphics[width=17.8cm]{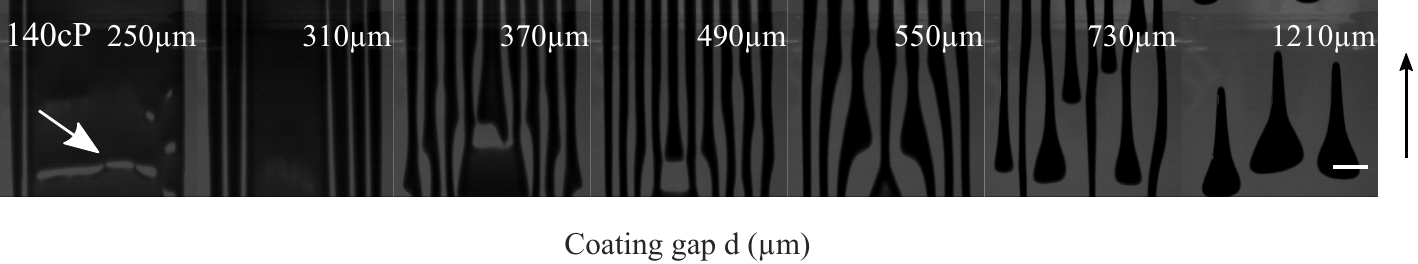}
    \caption{Transition from homogeneous coating to stripes upon increasing the coating gap from $d=\SI{250}{\micro\m}$ to $d=\SI{1210}{\micro\m}$ as given in each image with $v=\SI{0.7}{\meter\per\minute}$ for $\eta=\SI{140}{cP}$. Remaining details are as in Fig.~\ref{map}}\label{gap_140cP}
\end{figure*}

At 9\% PVP, which corresponds to $\eta=\SI{140}{cP}$, some elongated holes perpendicular to the coating direction appear in the otherwise homogeneous coating when increasing the coating gap height, see Fig.~\ref{gap_140cP}.
Note that here we vary the gap height $d$ for one set of parameters $v$ and $\eta$ where only
stripes perpendicular to the coating direction appear when increasing the coating gap height. However, these stripes are irregular and not as well defined as the stripes parallel to the coating direction which appear when increasing the coating gap height at the other viscosities.

Again the results may be converted to a dimensionless form when the gap height $d$ is rescaled by a characteristic reference length scale. We follow Ref.~\cite{CaKh2000aj} and compare the gap height to the film height of a homogeneous coating $h_\mathrm{hom} = q / (s \, v)$ for a given flow rate $q$ and slot width $s$. This results in a dimensionless gap height $\tilde d = d \, s \, v / q$ which ranges from $1.4$ - $16.1$ in the experiment depicted in Fig.~\ref{gap}.

\section{Theoretical approach}\label{sec:theo}
To compare the presented experimental results to theory we introduce a thin-film model for slot-die coating that takes capillarity and wettability into account as well as parameters of the coating process like coating velocity and gap height. We focus on the region close to the downstream meniscus, i.e., on the region directly trailing the moving coating die.
We consider a simple partially wetting nonvolatile liquid and base the model on a hydrodynamic long-wave or thin-film model~\cite{OrDB1997rmp,Thie2007,CrMa2009rmp}.  Such models are successfully employed to describe the meniscus dynamics and instabilities for related coating techniques, such as dip-coating~\cite{SADF2007jfm,ZiSE2009epjt,GTLT2014prl,GLFD2016jfm,TWGT2019prf}.

The evolution equation for the film thickness profile $h(\mathbf x, t)$ is obtained by a long-wave expansion of the Navier-Stokes equations in a laminar flow limit together with adequate boundary conditions at the solid substrate and a free surface~\cite{OrDB1997rmp,Thie2007}. Here, we adapt the resulting thin-film equation to the geometry of slot-die coating in the reference frame of the meniscus, i.e., moving with the coating die. In the dimensional form it reads
\begin{align}
    \frac{\partial}{\partial t} h(\mathbf x, t) &= -\nabla \cdot\mathbf j (\mathbf x, t)\nonumber\\
    &= -\nabla \cdot\left[ \frac{h^3}{3\eta} \nabla \left( \sigma \Delta h + \Pi(h) \right) \right] - v \ \partial_x h\label{eq:TFE}
\end{align}
where $\mathbf j$ is the liquid flux and the advection term with coating velocity $v$ accounts for the movement of the slot-die over the substrate. Moreover, $\Pi(h)$ is the Derjaguin or disjoining pressure that models partial wettability~\cite{DeG1985rmp,derjaguin1987surface,Thie2010jpcm}. We employ the common form
\begin{align}
    \Pi(h) = H_A \left( \frac{h_p^3}{h^6} - \frac{1}{h^3} \right).\label{eq:TFE_left_boundary}
\end{align}
that combines a destabilizing long-range van der Waals interaction and a short-range stabilizing interaction. Here, it is parametrized by the height of the equilibrium adsorption layer $h_p$ and the Hamaker constant $H_A = \frac{5}{3} \sigma h_p^2 \theta^2$ that we relate to the equilibrium contact angle $\theta$ employing Young's law~\cite{churaev1995contact}.
Since Eq.~\eqref{eq:TFE} describes the evolution of the height profile of the coated film, we limit the spatial domain to the red box in Fig.~\ref{SetUp}. That is, on the downstream side the position $x = 0$ marks the edge of the slot-die and gives the limiting point of the down-stream meniscus. The experimental gap height $d$ and imposed constant liquid influx $q$ then translate into the boundary conditions
\begin{align}
    h(x=0, t) = d \quad\text{and}\quad
    \mathbf{j}(x=0, t) = \frac{q}{s} \mathbf{\hat e}_x,
\end{align}
where $s$ is the width of the slot. We place the upstream boundary of the domain at a position $x=L$ far from the meniscus, and apply the Neumann condition
\begin{align}
    \frac{\partial}{\partial x} h(x=L, t) = \frac{\partial^3}{\partial x^3} h(x=L, t) = 0\label{eq:TFE_right_boundary}
\end{align}
that allows for free outflow of the liquid volume.

We rewrite the model in a dimensionless form. This prepares the equations for a numerical treatment but and provides insights into the scaling behavior of the solutions with respect to the parameters. Hence, we choose the gap height as a reference height scale $h_0=d$ and introduce the length scale $x_0 = \sqrt{3/5}\, d / \theta$ and time scale $t_0 = 27 \eta d / (25 \sigma \theta^4)$ so that the contact angle $\theta$, the viscosity $\eta$ and the surface tension $\sigma$ are eliminated from the governing equation:
\begin{align}
    \frac{\partial}{\partial \tilde t} \tilde h(\mathbf{\tilde x}, \tilde t) = &-\tilde \nabla \cdot\left\{ \tilde h^3 \tilde \nabla \left[\tilde \Delta \tilde h + \tilde h_p^2 \left( \frac{\tilde h_p^3}{\tilde h^6} - \frac{1}{\tilde h^3} \right) \right] \right\}\nonumber \\ &- \tilde{\mathrm{Ca}} \ \partial_{\tilde x} \tilde h.\label{eq:TFE_nondim}
\end{align}
Hereby, the number of free parameters is reduced to a scaled capillary number $\tilde{\mathrm{Ca}} = 27 \sqrt{5}\,\eta v/ (25 \sqrt{3}\,\sigma \theta^3)$, the dimensionless flow rate $\tilde q = q\ t_0 / (x_0 h_0)$ and the (typically small) ratio of the adsorption layer thickness to gap height $\tilde h_p = h_p / d$, that we fix at $\tilde h_p = 0.04$.

\begin{figure*}
    \centering
    \includegraphics[width=17.8cm]{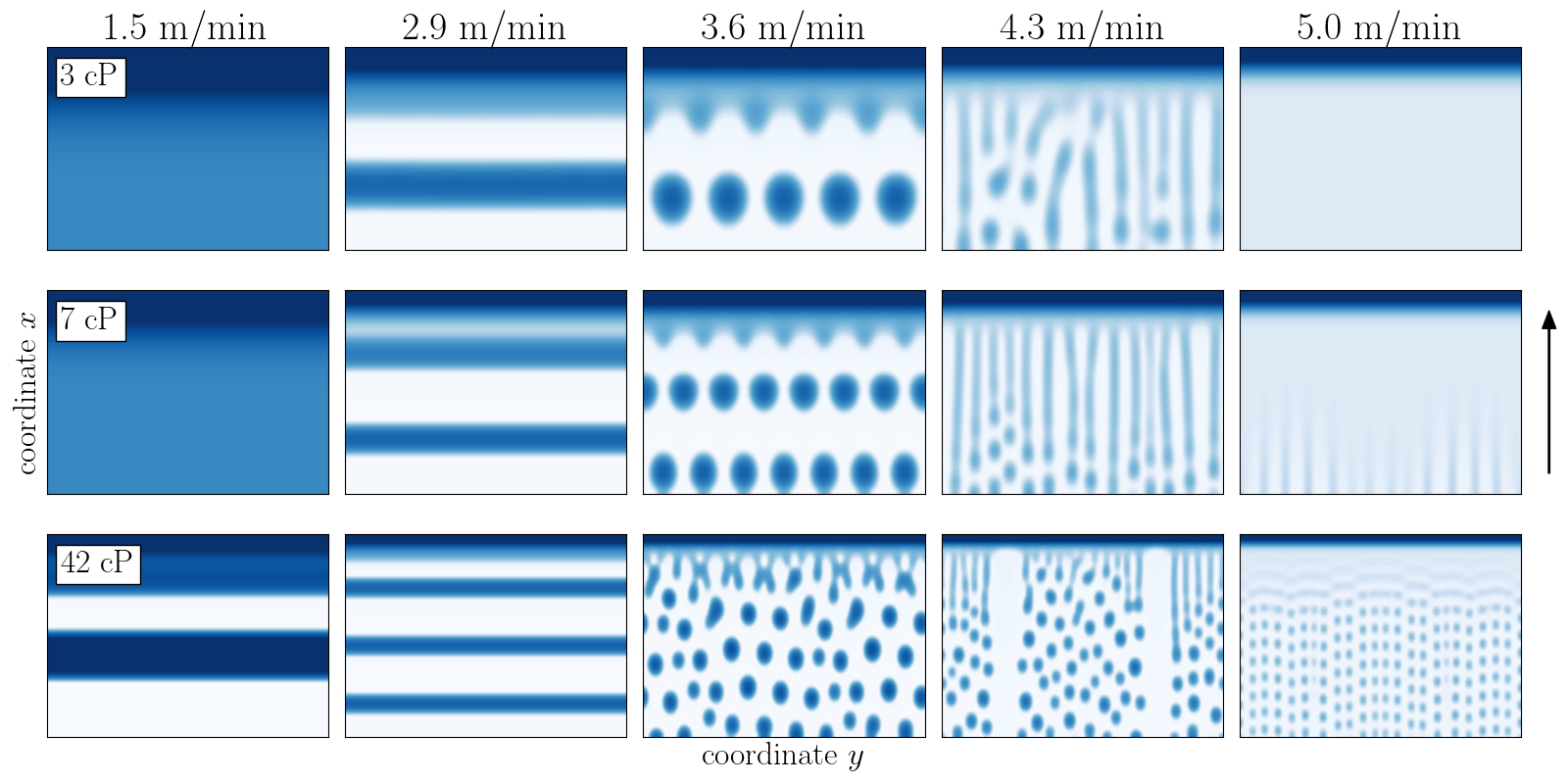}
    \caption{Snapshots of numerical simulations of the model Eq.~\eqref{eq:TFE} for various combinations of the viscosity and the coating speed after initial transients. Blue colors represent the liquid while uncoated areas are white. The slot is positioned at the top of the domain ($\SI{5}{\milli\m}\times\SI{2.5}{\milli\m}$) and moves upwards. Similar to the experimental results (cf. Fig.~\ref{map}) we find transitions between a homogeneous coating and different stripe and droplet patterns that strongly depend on the coating speed and viscosity. The arrow indicates the coating direction. Further details are discussed in the main text.}\label{fig:theo}
\end{figure*}

We employ direct numerical simulations of the model Eqs.~\eqref{eq:TFE}--\eqref{eq:TFE_right_boundary} using the finite element method (FEM) approach of the C++-library \textsc{oomph-lib}~\cite{HeHa2006} and a backward differentiation scheme of second order (BDF2) for time integration. The simulations are performed on a spatial domain of the size $\SI{5}{\milli\m}\times\SI{2.5}{\milli\m}$ with periodic boundaries in $y$-direction. Note that the periodic boundary conditions can cause defects in the coating patterns when the lateral domain size does not match the natural wavelength of a parallel stripe pattern.
Modeling our experimental setup, we fix the gap height at $d=\SI{220}{\micro\m}$, impose a flow rate $q=\SI{4.8}{\milli \litre / \minute}$ for a slot of width $s=\SI{8}{cm}$ and assume a surface tension of $\sigma = \SI{25}{\milli \newton/\m}$. Additionally, the contact angle is adapted to the viscosity by interpolating the data of the measurement presented in Fig.~\ref{CAvisc} (i.e. $\theta=\SI{17}{\degree}$ for $\eta=\SI{3}{cP}$, $\theta=\SI{23}{\degree}$ for $\eta=\SI{7}{cP}$ and $\theta=\SI{38}{\degree}$ for $\eta=\SI{42}{cP}$). The simulation is initialized with a static $\tanh$-shaped meniscus that interpolates between the boundary condition $h(x=0)=d$ and the `dry' adsorption layer $h(x=L)=h_p$ plus a weak modulation of long wavelength. After numerically integrating the governing equations until transient effects have disappeared, we observe homogeneous coating layers as well as the formation of various stripe and droplet patterns at the meniscus. In analogy to Fig.~\ref{map}, we show in Fig.~\ref{fig:theo} the transition between these patterns for representative configurations of the coating speed and the viscosity.

In agreement with the experiment, at low velocities of the coating die we obtain homogeneously coated films (e.g.~see $v=\SI{1.5}{\meter\per\minute}$ in Fig.~\ref{fig:theo}). The height of the coated layer far from the meniscus is $h_\mathrm{hom} = q/(s v)$, as expected due to the conservation of liquid volume. As in the experiments, in the calculations this homogeneous coating becomes unstable when the velocity increases above a critical value, i.e., when the coating height $h_\mathrm{hom}$ decreases below  a critical value. Similarly, the same instability is observed at lower coating speeds when the viscosity becomes large, since both the viscosity and the coating speed enter the capillary number in the dimensionless dynamics (Eq.~\ref{eq:TFE_nondim}). Beyond the instability threshold a stripe pattern emerges, that is oriented perpendicular to the coating direction. In the column with $v=\SI{2.9}{\meter\per\minute}$ in Fig.~\ref{fig:theo} we observe that the wavelength of the pattern decreases with increasing viscosity. This is due to the change of the contact angle with the viscosity and therefore due to a change in the length scale $x_0$ in the dimensionless dynamics.

When the coating speed is further increased to $v=\SI{3.6}{\meter\per\minute}$, the meniscus shows an additional transversal instability that leads to the periodic formation of droplets instead of stripes, resembling, e.g.,~the transition shown in the top row of Fig.~\ref{map}(b) where droplets are found when the velocity exceeds $v_\mathrm{crit,irregular}$. Note that in the simulations we see no significant dependence on viscosity in the critical coating speed of this transition. As for the stripe pattern, viscosity influences the wavelengths of the drop patterns. In the vicinity of the transition, the system may show long transients or intermittent/mixed behavior where bursts of drops and stripes can be observed concurrently. In this regime, finite size effects and perturbations may play an important role in the selection of the pattern.

At even larger coating speeds, fingers of liquid can grow from the meniscus and thus form a stripe pattern that is oriented parallel to the coating direction (e.g.~see $v=\SI{4.3}{\meter\per\minute}$). Note that these stripes will break up into single droplets if they become too long and also defects may be induced if the lateral width of the (periodic) domain is incompatible with the wavelength. Regular stable regimes exist for selected domain sizes. Again, the wavelength is observed to scale with the viscosity as indicated by the dimensional analysis above. If the wavelength of the parallel stripes is very small (e.g.\ for $\eta=\SI{42}{cP}$), the pattern is interrupted by broad uncoated areas while the meniscus confines the liquid into smaller regions, preserving the pattern. Again, transient/intermittent regimes between parallel stripes and droplet patterns can be found near the critical parameter values.

For relatively large velocities (e.g. $v=\SI{5.0}{\meter\per\minute}$), the effective coating layer height $q/(s v)$ is no longer large as compared to the adsorption layer height $h_p$. In this regime, the model reveals ultra-thin homogeneous coatings that are stable on small domains but show dewetting behavior if the domain is large enough compared to the wavelength of the dewetting instability (see~\cite{Thie2007}). Accordingly, we see a stable layer for $\eta=\SI{3}{cP}$, a lateral dewetting instability for $\eta=\SI{7}{cP}$ and dewetting towards droplets for $\eta=\SI{42}{cP}$. We stress that these patterns do not result from an instability at the meniscus and depend on the choice of the wetting potential.

We also note that in the simulation we always first find a transition from a homogeneous film to perpendicular stripes while in the experiment this only occurs at large viscosities. Furthermore, there is a slight quantitative mismatch in the scales of the simulation and the experiments. Therefore, in Fig.~\ref{fig:theo} we have chosen the parameters such that a good overview is given of the patterns that can be observed in the simple thin-film  model. The discussion of these deviations is continued in the next section.

\section{Discussion}\label{sec:diss}
We have analyzed the occurrence of patterned coatings in slot-die coating of thin layers of PVP-ethanol mixtures. The mass concentration of PVP has been employed to vary the viscosity of the mixture, which also changes the contact angle of the liquid. The coating velocity and coating gap height have been used as further main control parameters. As a result we have found that above a critical coating velocity the uniform coating becomes unstable with respect to patterning. At high viscosities we have found stripe patterns oriented perpendicular to the coating direction while at low viscosities stripes oriented in parallel direction appear. At intermediate viscosities coexistence of both types of stripes was observed. Further increasing the velocity, in all three cases the patterns acquire more and more defects and finally become rather irregular. Additionally, pattern formation was also observed when increasing the coating gap height. In this case, parallel stripes appear regardless of the fluid viscosity. We found that the period of the stripe pattern decreases when the coating gap height is further increased.

These findings consistently expand results known from the literature. In particular, Raupp et al.~\cite{Raupp2018} describe the formation of parallel stripes in a range of coating velocities as rivulets. The also described ribbing -- a weaker modulation of film thickness also parallel to the coating direction -- has not been found here. Schmitt et al.~\cite{Schmitt} find
perpendicular stripes at high viscosities with parameter dependencies that qualitatively agree with our findings. Parallel and perpendicular stripe formation is also observed in further studies~\citep{khandavalli2016, kang2014}. However, the coexistence of stripes in the two different directions has to our knowledge not yet been reported in the literature.

We have also presented results for the dependency of the period of stripe patterns on coating velocity. Overall, our experiments indicate that with increasing coating velocity the inverse period (wavenumber) first increases before it decreases again. Note that the result is not fully conclusive as the qualitative behavior seems to change with viscosity. On the basis of the obtained data one cannot yet deduce the exact bifurcation character of the onset of pattern formation. Overall, the behavior of the period resembles the one observed in Langmuir-Blodgett transfer where it is also found that the wavenumber first increases and then decreases with transfer velocity~\cite{LKGF2012s}. Qualitatively similar behavior is also observed with models of solute deposition by dip coating solutions with volatile solvent~\cite{DoGu2013e,DeDG2016epje}.

Beside the experimental results, we have also presented a simple thin-film model describing the dynamics of the meniscus and deposited film. It incorporates capillarity and wettability and has allowed us to obtain a typical sequence of patterns as found when increasing the coating speed, namely, the transition from uniform coating to perpendicular stripes, droplets and parallel stripes. However, the agreement is only partial, since there are both, qualitative and quantitative differences. Qualitatively, the thin-film model only reproduces part of the sequences of the various pattern types seen in the experiment. There exists, e.g.,~a direct transition from uniform layers to parallel stripes while the model only gives transitions from a uniform layer to perpendicular stripes. However, we emphasize that the model is able to reproduce the different experimentally observed patterns on comparable scales. A future in-depth bifurcation study could give further insights into all possible sequences of patterning including possible multistabilities of different pattern types.

Notably, the model does not show a strong dependence of the dominant pattern on viscosity. If viscosity and coating velocity both only entered the capillary number Ca, all relevant borders in the morphological phase diagram Fig.~\ref{map} would be hyperbolas $\eta\sim v^{-1}$. This is not the case, a fit of the onset of patterning in Fig.~\ref{map} actually gives $\eta\sim v^{-1/2}$, i.e., there must be other relevant nondimensional numbers beside the capillary number. Based on the experimental results of the dependency of contact angle on PVP concentration, we incorporated the influence of wettability into the modeling. This results in some influence on the observed patterns but less strong than in the experiments.
For a deeper analysis of the influences of viscosity and wettability in future investigations, it would be helpful to find experimental systems where the two can be independently changed.

In the model, we have found that wettability, in particular, the height of the adsorption layer, is critical for achieving a quantitative comparability of the observed length scales. We expect that the experimental and theoretical scales can be further aligned, if the wettability was independent from the viscosity in the experiments or the effects underlying the observed dependence were better understood and incorporated into the model. Further improvement could include a discussion of the influence of non-Newtonian effects in all stages of pattern formation, in particular, when local shear rates strongly deviate from the averaged values our above estimates rely on. This may include shear-thinning, stretching force and viscoelasticity of the polymer solution as observed, e.g., for liquid jets~\cite{yarin1993free,Kolbasov2015}.

\begin{figure}[tbh]
    \centering
    \includegraphics[width=8.6cm]{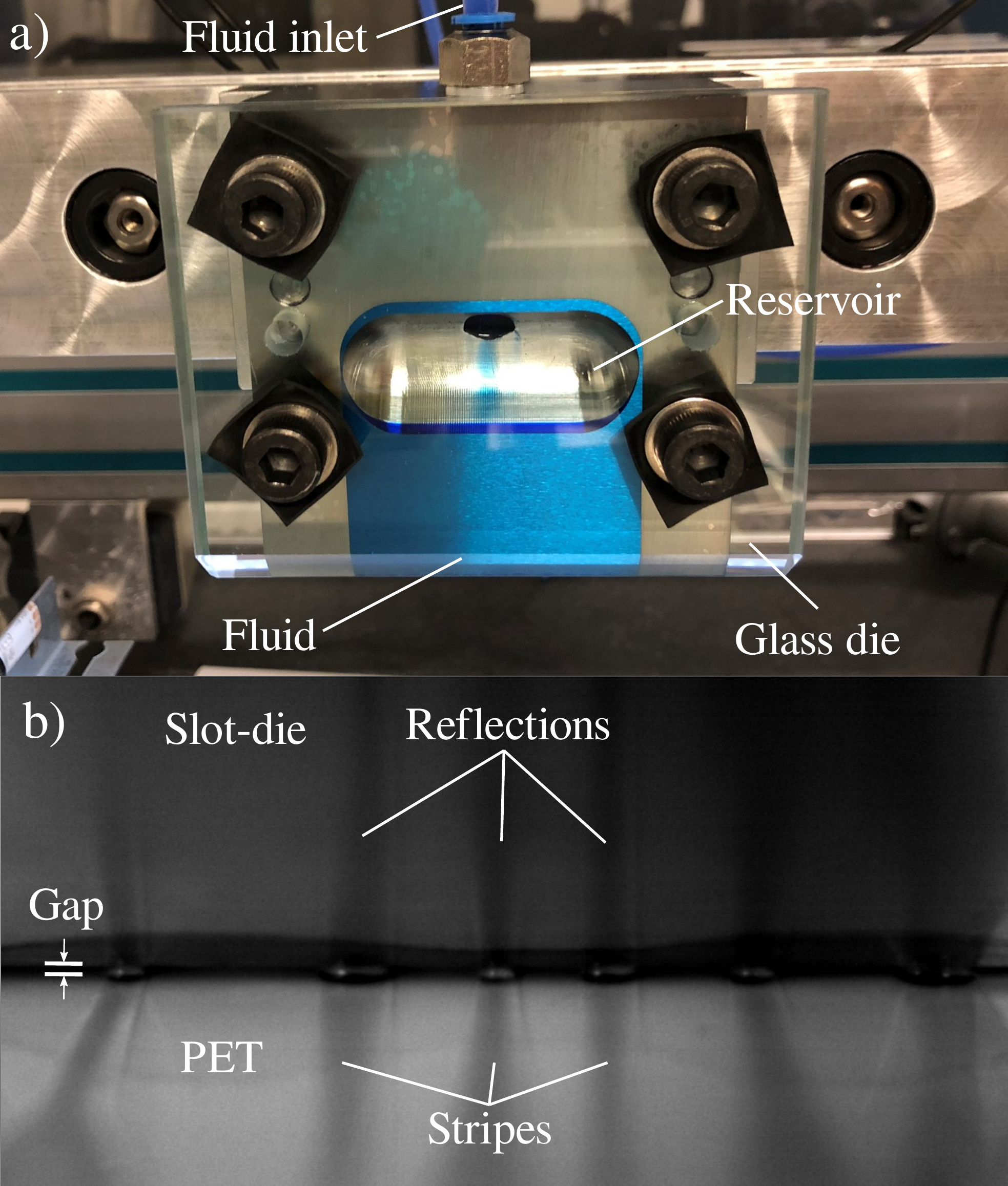}
    \caption{(a) Image of the slot-die system with glass front plate used to control the coating liquid homogeneity in the die. (b)~Image made during the coating with a CCD camera installed on the slot-die: the coating gap is in the middle; the stripes on PET (also seen in the reflection on the slot-die) are formed immediately at the die.}\label{Hom2Stripe}
\end{figure}

The research of Carvalho and Kheshgi~\cite{CaKh2000aj} and Romero et al.~\cite{RSSC2004jnfm}, suggests that coating defects may occur in the so-called ``low-flow limit'' when the curvature of the downstream meniscus increases until it fails to bridge the coating gap height. Since we have only modeled the height profile of the downstream meniscus, this effect is excluded in our theoretical description. To incorporate this mechanism into the model, one would need to include the hydrodynamics within the gap. Another discussed effect is the entrainment of air bubbles at the upstream meniscus in Fig.~\ref{SetUp}~(d). The bubbles are then transported underneath the die to the upstream meniscus. However, this normally results in air bubbles enclosed in the coating, not the coating patterns studied here~\cite{Bhamidipati2012}.

\modified{The importance of the flow under the die is confirmed by a further observation: 
to investigate the influence of solvent evaporation, a video camera has been installed directly on the die. White diodes illuminate the coating gap from the backside and pattern formation can be observed directly in the coating gap of 
\SI{220}{\micro\m} height between the die and the substrate.} These additional 
experiments have clearly demonstrated that the parallel stripes appear directly 
at the die and are neither due to a later rearrangement of the liquid on the 
substrate nor do they result from solvent evaporation. A snapshot of the die 
during the coating process is provided in Fig.~\ref{Hom2Stripe}~(b).

\modified{To analytically assess the role of evaporation and potential Marangoni flows during the coating process, we consider the Marangoni number $\mathrm{Ma}=-\frac{\partial \sigma}{\partial T} \frac{H \Delta T}{\eta \alpha}$ where $\Delta T$ is the vertical temperature difference in the coated layer of characteristic height $H$ and $\alpha=\SI{0.08}{mm^2/s}$ is the thermal diffusivity of ethanol~\cite{Lide2004}. To estimate $\mathrm{Ma}$, we assume that the dependency of the surface tension on the temperature is of order -\SI{0.3}{mN/(m\,K)} (similar to aqueous solutions~\cite{AgTG2011jced}), and that the evaporation-induced temperature difference is $\Delta T = \SI{0.25}{K}$ (as e.g.~in Ref.~\cite{ToSS2021p}). Further, we use a characteristic film height of $H=\SI{20}{\micro m}$. Then, the Marangoni number is in the range $\mathrm{Ma} \in [0.02, 0.63]$ for viscosities $\eta \in [\SI{10}{cP}, \SI{400}{cP}]$, supporting our assumption that thermal Marangoni effects are insignificant during the coating process. Moreover, solutal Marangoni flows cannot explain the pattern formation, as the surface tension depends weakly on the PVP concentration and, here, an evaporation of the solvent, i.e.\ an accumulation of PVP at the surface, would actually induce a stabilizing flow.
However, we note that the later occurring drying of the coated patterned layers indeed results in an additional low-amplitude small-scale pattern. Due to the rather weak Marangoni flows~\cite{ChDC2012e}, it only appears several minutes after the coating process has ended. The resulting secondary pattern is of a different (convective) nature than the primary pattern discussed in our work. It is beyond our present scope.}

Further, it has been checked that the parallel stripes are not already produced by flow instabilities within the die before the liquid leaves the slot. To do so, the metallic front plate of the die has been substituted by a transparent glass plate, see Fig.~\ref{Hom2Stripe} (a). Additional experiments (not shown here) have confirmed (i) that the replacement of metal by glass does not noticeably change the observed parameter ranges and the transition scenarios, and (ii) that the flow inside the die is always homogeneous (also no air bubbles were observed in the slot channel). 

In both, experiment and simulations, the patterns were susceptible to defects.
To acquire an understanding of the topology of defects and the stability of the patterns to perturbations, the experimental setup needs to be extended in order to allow for longer coated areas.

In conclusion, focusing on parameter regimes outside the coating window, we have achieved some insight into the formation of patterns in slot-die coating and their sequence of occurrence. Changing several principal parameters, we have revealed transitions from homogeneous coating inside the coating window to various stripe patterns, mixed and irregular patterns. Furthermore, we have established a relatively simple thin-film model for the coating hydrodynamics at the meniscus. In this way we have been able to model an identical set of patterns and pertinent transitions. A discussion of agreements and disagreements of our experimental and theoretical approaches has let us to outline possible steps for further research.
\begin{acknowledgments}
The authors acknowledge the financial support by the German Science Foundation (DFG) through grants TH781/8-1 and GU1075/14-1.
\end{acknowledgments}

\bibliography{CoatingPaper}

\end{document}